\documentclass[%
prl%
,reprint%
,twocolumn%
,secnumarabic%
,floatfix%
,superscriptaddress
,showpacs%
,amssymb, amsmath, nobibnotes, aps]{revtex4-1}

\usepackage[latin1]{inputenc}
\usepackage{epsfig}
\usepackage{color}

\newcommand{\ind}[1]{_{\text{#1}}}
\newcommand{\bra}[1]{\left<#1\right|}
\newcommand{\ket}[1]{\left|#1\right>}

\begin{document}

\title{Collective excitation of Rydberg-atom ensembles beyond the $\sqrt{N}$ enhancement}

\author{Martin G\"arttner}
\affiliation{Max-Planck-Institut f\"{u}r Kernphysik, Saupfercheckweg 1, 69117 Heidelberg, Germany}
\author{Shannon Whitlock}
\affiliation{Physikalisches Institut, Universit\"at Heidelberg, Im Neuenheimerfeld 226, 69120 Heidelberg, Germany}
\author{David W.\ Sch\"{o}nleber}
\affiliation{Max-Planck-Institut f\"{u}r Physik komplexer Systeme, N\"othnitzer Stra{\ss}e 38, 01187 Dresden, Germany}
\affiliation{Max-Planck-Institut f\"{u}r Kernphysik, Saupfercheckweg 1, 69117 Heidelberg, Germany}
\author{J\"org Evers}
\affiliation{Max-Planck-Institut f\"{u}r Kernphysik, Saupfercheckweg 1, 69117 Heidelberg, Germany}

\date{\today}

\begin{abstract}
In an ensemble of laser-driven atoms involving strongly interacting Rydberg states, the  excitation probability is usually strongly suppressed. In contrast, here we identify a regime in which the steady-state Rydberg excited fraction is enhanced by the interaction. This effect is associated with the build-up of many-body coherences, induced by coherent multi-photon excitations between collective states. The excitation enhancement  should be observable under currently-existing experimental conditions, and may serve as a direct probe for the presence of coherent multi-photon dynamics involving collective quantum states.
\end{abstract}

\maketitle

The emergence of collective quantum effects in a many-body system is a hallmark of the strongly-correlated regime. A celebrated early example is the phenomenon of Dicke superradiance, in which $N$ two-level atoms coherently interacting with a common optical field acquire enhanced emission properties \cite{dicke1954}. In cavity quantum electrodynamics, an ensemble of $N$ atoms placed inside an optical cavity experiences a collective enhancement of the couping to a single cavity photon which scales as $\sqrt{N}$ \cite{colombe2007, brennecke2007}. Similar collective effects arise in the coupling of superconducting devices to nitrogen vacancy centers in diamond \cite{kubo2010, zhu2011, amsuess2011}, and could play important roles in circuit-QED \cite{wallraff2004} and hybrid optical-micromechanical systems \cite{vogell2013}. Ensembles of highly-excited Rydberg atoms offer an alternative system with which to study correlations and collective effects due to strong interatomic interactions. Here the competition between the laser excitation process and the interactions ensures only one Rydberg excitation can be accommodated within a critical distance (dipole blockade effect) leading to strong spatial correlations. In this regime a $\sqrt{N}$ enhancement of the atom-light coupling 
has been demonstrated~\cite{heidemann2007, gaetan2009, urban2009, dudin2012c}. By exploiting these effects it should be possible to realize new quantum technologies such as non-classical light sources \cite{dudin2012b, peyronel2012, maxwell2013} and quantum gates based on collectively enhanced interactions~\cite{lukin2001}. 

Here, we point out a surprising enhancement of the steady-state Rydberg population for driven three-level atoms which goes beyond the $\sqrt{N}$ enhancement typically associated with the dipole blockade effect. The enhancement occurs for resonant two-step excitation from the ground state to the Rydberg state (under electromagnetically-induced-transparency (EIT) conditions) and for repulsive Rydberg-Rydberg interactions. We show that the effect is related to direct multi-photon transitions between collective states. As a result, its investigation requires exact calculations of the many-body master equation or analytical calculations beyond second order in the driving fields. We identify regimes in which the effect is most pronounced, and find that the Rydberg population of the repulsively interacting system can even be larger than for the completely non-interacting system, highlighting the genuine many-body nature of the system. As this effect should be observable in existing cold atom experiments, it may serve as a direct probe for coherent multi-photon processes between collective states via the observation of a global steady state observable.

\begin{figure}[t]
  \centering
 \includegraphics[width=\columnwidth]{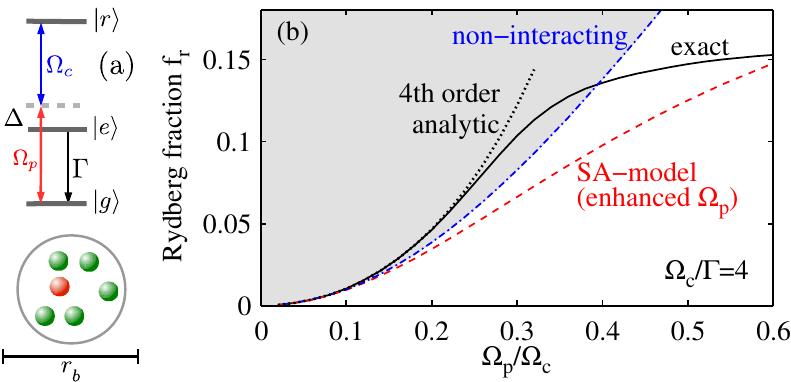}
 \caption{(Color online) (a) Illustration of the setup: Three-level atoms with the Rydberg state $\ket{r}$. The blockade radius exceeds the trap size such that at most one atom can be in the Rydberg state. (b) Steady-state values of the Rydberg fraction $f_r$ for $N=4$ atoms as a function of the probe Rabi frequency ($\Delta=0$). 
 The coherent laser driving is strong compared to the decay rate $\Gamma$ of the intermediate level. Solid black line: exact ME solution, red dashed line: result of a SA model with $\sqrt{N}$-enhanced $\Omega_p$, blue dot-dashed line: value of $f_r$ for non-interacting atoms, black dotted line: analytical solution for $f_r$ in the weak probe limit. The gray shading denotes the region of interaction-induced excitation enhancement beyond the non-interacting value.}
 \label{fig:MERE_pop}
\end{figure}

First, we consider an ensemble of $N$ three-level atoms all confined to a volume comparable to, or smaller than, a single blockade sphere (Fig.~\ref{fig:MERE_pop}a). The atoms are driven by two laser fields which couple the ground and intermediate state $\ket{g}\rightarrow \ket{e}$ and the intermediate and the Rydberg state $\ket{e}\rightarrow \ket{r}$, referred to as the probe and coupling transitions, respectively. The $\ket{e}$ state decays rapidly (decay rate $\Gamma$) via spontaneous emission to the ground state while the other two states are long-lived. 
 
The Hamiltonian of this system reads ($\hbar=1$)    
\begin{equation}
H=\sum_{i=1}^N H_L^{(i)} - \sum_{i=1}^N \Delta\ket{e_i}\bra{e_i} + \sum_{i<j}\frac{C_6 \ket{r_i r_j}\bra{r_i r_j}}{|\mathbf{x}_i-\mathbf{x}_j|^6}
 \label{eq:Hamiltonian}
\end{equation}
where $H_L^{(i)}=\Omega_p/2 \ket{g_i}\bra{e_i} + \Omega_c/2 \ket{e_i}\bra{r_i} + h.c.$ and $\mathbf{x}_i$ are the atomic positions. We allow for a detuning $\Delta$ from the intermediate state while the two-photon transition is kept resonant. We assume that the Rydberg states interact repulsively via isotropic van der Waals interactions with strength $C_6$. Incoherent processes like spontaneous emission and dephasing are treated using a master equation (ME) including Lindblad terms \cite{supplement}. The resulting ME reads $\dot{\rho}=-i[H,\rho]+\mathcal{L}[\rho]$.
We assume continuous, spatially homogeneous laser driving and focus on the Rydberg fraction $f_r=\mathrm{Tr}[(N^{-1}\sum_i \ket{r_i}\bra{r_i}) \rho]$ in the steady state ($\dot{\rho}=0$) as our main observable. 
We are able to simulate the dynamics and steady states for up to $N=10$ three-level atoms using the wave function Monte Carlo method \cite{molmer1993}. 

Solving the ME for a single atom ($N=1$) and under perfect EIT conditions (zero dephasing and decay of the Rydberg state), up to a normalization factor, yields the steady state $\ket{d}=\ket{g}-\Omega_p/\Omega_c\ket{r}$, which is the EIT dark state. Thus the Rydberg excitation probability, equivalent to the steady state Rydberg fraction, becomes $f_0=\Omega_p^2/(\Omega_c^2+\Omega_p^2)$. 
In the case of a fully blockaded ensemble of $N>1$ particles, the atoms are excited to collective Dicke states, which leads to a $\sqrt{N}$ enhancement of the atom-light coupling \cite{dicke1954,lukin2001}. In this spirit one can describe the fully blockaded ensemble as a single  three-level ``super-atom'' (SA) with a $\sqrt{N}$ larger transition dipole moment for the probe transition. The resulting Rydberg excitation probability of a SA is obtained by replacing $\Omega_p$ with $\sqrt{N}\Omega_p$ in $f_0$, i.e., $N f\ind{SA}=N\Omega_p^2/(\Omega_c^2+N\Omega_p^2)$. Comparing $f\ind{SA}$ to $f_0$ shows that in the simple SA picture, the SA excitation probability cannot be enhanced more than $N$-fold over the single-atom value,  $Nf\ind{SA}\leq Nf_0$. In the following we will show that the exact solution of the ME can violate this bound and thus an excitation enhancement beyond the bare $\sqrt{N}$ is possible. 

Figure~\ref{fig:MERE_pop}(b) shows the Rydberg fraction $f_r$ for a fully blockaded ensemble as a function of $\Omega_p/\Omega_c$ for $N=4$, $\Omega_c = 4\Gamma$ and $\Delta=0$. For weak driving $\Omega_p\rightarrow 0$ the exact solution agrees with the non-interacting (or single-atom) value $f_0$. For large $\Omega_p$ the Rydberg fraction $f_r$ approaches a constant value of $\approx 1/N$ due to the blockade. In between (for $0.15 \lesssim\Omega_p/\Omega_c\lesssim 0.4$), the Rydberg excited fraction significantly exceeds the non-interacting value. This feature is counter to the usual expectation for the dipole blockade in which repulsive interactions lead to a reduction of the number of Rydberg excitations compared to the non-interacting case. The red dashed line shows the prediction of the SA model, which predicts $f\ind{SA}\leq f_0$ and is thus insufficient to explain this feature.

\begin{figure}[t]
  \centering
 \includegraphics[width=\columnwidth]{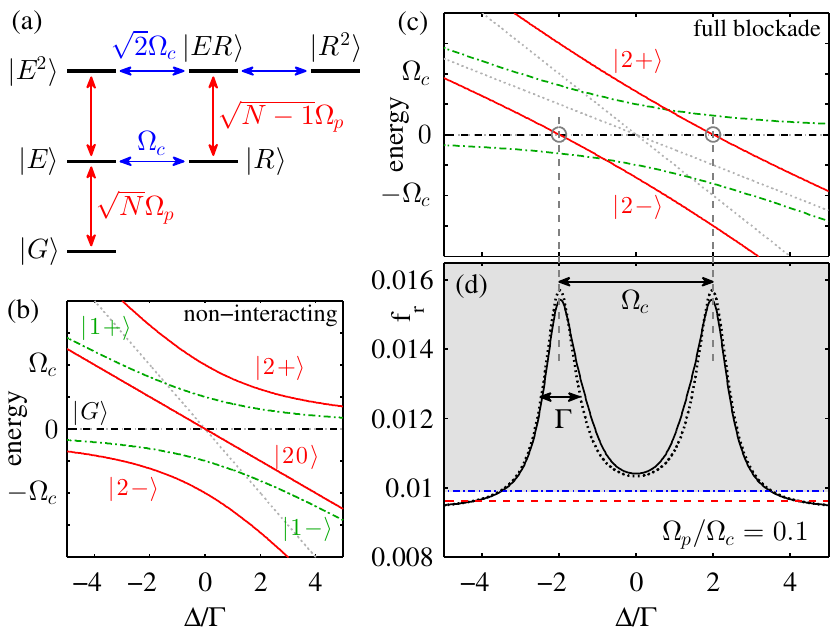}
 \caption{(Color online) (a) Level scheme of symmetrized (Dicke) states. Diagonalizing the horizontal couplings ($\Omega_c$) a dressed state picture is obtained. (b) and (c) Dressed state energies as a function of the detuning from the intermediate state $\Delta$ in the (b) non-interacting and (c) blockaded case. In the blockaded case resonances between the state $\ket{G}$ and the doubly excited dressed states (red) occur at $\Delta\approx \pm\Omega_c/2$ (circles). Gray dotted lines show bare energies (or asymptotes). (d) Rydberg fraction as a function of $\Delta$. The parameters are as in Fig.~\ref{fig:MERE_pop}. Solid black line: exact ME solution, red dashed line: result of the SA model, blue dot-dashed line $f_0$, black dotted line: analytical model. The resonances between the dressed Dicke states result in a large excitation enhancement in comparison to the non-interacting Rydberg fraction.}
 \label{fig:delta_dep}
\end{figure}

To provide a physical picture for the collective excitation enhancement we express the system in the basis of symmetrized Dicke states \cite{dicke1954, lukin2001, carmele2013, liu2014}. These states are fully symmetric superpositions of all excited states with the same number of $e$ and $r$ excitations,
\begin{equation}
 \ket{E^j R^s}=\mathcal{N}_{j,s}\left(\sum_{i=1}^N\ket{e_i}\bra{g_i}\right)^j\left(\sum_{i=1}^N\ket{r_i}\bra{g_i}\right)^s\ket{G}
\end{equation}
with the normalization $\mathcal{N}_{j,s}$ \cite{supplement}. We use the shorthand notation $\ket{E^0 R^0}=\ket{G}$, $\ket{E^1 R^0}=\ket{E}$ and $\ket{E^0 R^1}=\ket{R}$ [Fig.~\ref{fig:delta_dep}(a)]. Note that $s\in\{0,1\}$ for a perfectly blockaded ensemble, so the state $\ket{R^2}$ is not present in this case. The Hamiltonian \eqref{eq:Hamiltonian} preserves the symmetry of these states and population of states outside the manifold of symmetric states only arises due to incoherent effects such as spontaneous emission. The coupling laser does not change the total number of excitations $j+s$. Therefore, the $\Omega_c$-part of the Hamiltonian can be diagonalized, resulting in the dressed many body eigenstates shown in Figs.~\ref{fig:delta_dep}(b) and (c). 
Due to the exclusion of states with multiple Rydberg excitations the interacting system exhibits new level crossings. In particular we point out the crossings between the states $\ket{G}$ and the dressed states of the $j+s=2$ manifold at $\Delta=\pm\Omega_c/2$ where we expect an excitation enhancement due to direct two-photon processes. In the non-interacting case, a level crossing appears at $\Delta=0$, however, here the dressed Dicke state picture is incomplete since interference effects between different excitation channels are not visualized. Such effects lead to the absence of $e$-excitations in the non-interacting case (EIT) and thus the non-interacting Rydberg fraction is independent of the intermediate state detuning $\Delta$.
Figure~\ref{fig:delta_dep}(d) shows the detuning dependence of the Rydberg fraction. The full calculations show a significant detuning dependence of $f_r$ with two peaks centered around $\Delta\approx \pm\Omega_c/2$ with widths of $\approx \Gamma$. At its maximum the enhancement is approximately a 50\% effect. On resonance ($\Delta=0$) the enhancement is still visible, but only on the order of 5\% for these parameters. Clearly, here the simple SA picture fails, since it amounts to neglecting all but the states $\ket{G}$, $\ket{E}$, and $\ket{R}$. The presence of the resonances, however, shows that states with $j+s>1$ play a crucial role. In particular, from the dressed state representation we find that coherent two-photon excitation from the ground to the doubly excited collective states is essential for the observed enhancement effect.

\begin{figure}[t]
  \centering
 \includegraphics[width=\columnwidth]{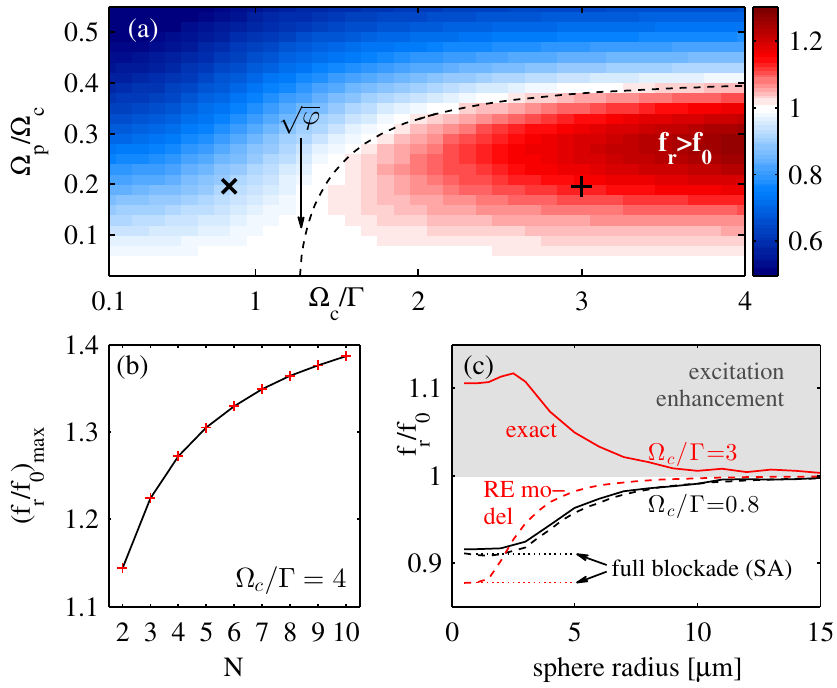}
 \caption{(Color online) (a) Dissipative phase diagram showing the enhancement factor $f_r/f_0$ as a function of Rabi frequencies for $N=4$ atoms and $\Delta=0$. For strong coherent driving and weak probe field, a region of $f_r>f_0$ is encountered. The dashed line shows the transition from suppressed to enhanced excitation ($f_r=f_0$). The arrow marks the critical value of $\Omega_c/\Gamma \approx 1.27$. (b) Maximal value of $f_r/f_0$ as a function of atom number $N$. (c) Few atoms ($N=4$) in a spherical volume of varying radius with random position sampling. The possibility for multiple Rydberg excitations is included in the simulation and realistic values for the laser dephasings and the spontaneous decay rate of the Rydberg state have been used (see text for details). The parameters correspond to the the black cross (lower black curves) and plus sign (upper red curves) in (a).  For the higher Rabi frequencies the excitation enhancement is present at all sphere radii (densities). }
 \label{fig:fbl_ME}
\end{figure}

On resonance, this effect appears most surprising as in the non-interacting case the dressed Dicke states show a crossing at $\Delta=0$, while in the blockaded case they do not.
Therefore,  in the following we focus on the excitation enhancement for the special case of $\Delta=0$.  Figure~\ref{fig:fbl_ME}(a) shows the ratio $f_r/f_0$, obtained from exact ME calculations as a function of the Rabi frequencies for the case of four fully blockaded atoms and perfect EIT conditions. For strong coupling and weak probe fields we observe a range of parameters where the Rydberg fraction exceeds that of a non-interacting ensemble, $f_r>f_0$. The dashed line marks the border of this region ($f_r=f_0$). As $N$ is increased the qualitative features of this dissipative phase diagram stay the same and the structure is compressed vertically, consistent with a $\sqrt{N}$ rescaling of $\Omega_p$, while the onset in $\Omega_c/\Gamma$ is independent of $N$ in the weak probe limit. We also determine the values of $\Omega_p$ for which $f_r/f_0$ is maximized. For $\Omega_c=4\Gamma$ the maximum is found at $\sqrt{N}\Omega_p\approx0.5\Omega_c$. Figure~\ref{fig:fbl_ME}(b) shows the maximal value of $f_r/f_0$ as a function of the atom number $N$. We find that $f_r/f_0$ appears to saturate but reaches values larger than $1.4$ if extrapolated to larger $N$. Similarly, the enhancement factor saturates as a function of $\Omega_c/\Gamma$.

In the following we derive analytical expressions for the Rydberg fraction for a fully blockaded ensemble using fourth order perturbation theory for small $\Omega_p/\Omega_c$. By dividing the ME by $\Omega_c/2$, we separate off the term associated with probe laser proportional to $\epsilon=\Omega_p/\Omega_c$ which can be treated as a perturbation in the weak probe limit. We then expand the steady state density matrix in orders of the small parameter $\epsilon$ ($\rho=\sum_n \rho_n \epsilon^n$) and solve for the $\rho_n$ recursively \cite{supplement}. To fourth order we obtain $f_r^{(4)} = \epsilon^2 + c_4\epsilon^4 + \mathcal{O}(\epsilon^6)$ where
\begin{equation}
\label{eq:fr4main}
  c_4  = 2(N-1)\frac{1+4\mathrm{Im[\beta]}^2-|\beta|^2-|\beta|^4}{|1+\beta^2|^2}-1
\end{equation}
with $\beta=(\Gamma+2i\Delta)/\Omega_c$. The predictions of Eq.~\eqref{eq:fr4main} are shown as dotted lines in Figs.~\ref{fig:MERE_pop}(b) and \ref{fig:delta_dep}(d).
We can now compare the fourth order terms in the non-interaction case $f_0=\epsilon^2-\epsilon^4+\mathcal{O}(\epsilon^6)$, the SA case $f\ind{SA}=\epsilon^2-N\epsilon^4+\mathcal{O}(\epsilon^6)$, and the master equation calculation $f_r^{(4)}$. By solving $c_4>-1$ for $\beta$ with $\Delta=0$ we find that an enhancement ($f_r>f_0$) occurs if $\Omega_c/\Gamma>\sqrt{\varphi}$ with $\varphi=(\sqrt{5}-1)/2$, independent of $N$ [arrow in Fig.~\ref{fig:fbl_ME}(a)]. A closer analysis of $c_4$ reveals that in the limit $\Omega_c\gg \Gamma$, Eq.~\eqref{eq:fr4main} describes two Lorentzian peaks of width $\Gamma$ centered at $\Delta=\pm\Omega_c/2$. The enhancement of the fourth order term beyond the non-interacting value is $c_4+1\approx 2(N-1)$ on resonance ($\Delta=0$), while at the maxima ($\Delta=\pm\Omega_c/2$) one obtains $c_4+1\approx 3\Omega_c^2(N-1)/(2\Gamma^2)$.

We now ask if the excitation enhancement persists under realistic experimental conditions, for example with imperfect blockade and including dephasing and the finite lifetime of the Rydberg state. We consider $^{87}$Rb atoms with states $\ket{g}=\ket{5s_{1/2}}$, $\ket{e}=\ket{5p_{3/2}}$, and $\ket{r}=\ket{55s}$ as in \cite{hofmann2012}. The atoms are assumed to be randomly distributed inside a sphere of variable radius. The corresponding interaction coefficient is $C_6/2\pi=50\,\textrm{GHz}~\mu\textrm{m}^6$, and the spontaneous decay rates are $\Gamma/2\pi=6.06$\,MHz and $\Gamma_r/2\pi=2$\,kHz. An overview of the simulations is shown in Fig.~\ref{fig:fbl_ME}(c) (solid lines). The lower black lines correspond to the Rabi frequencies $\Omega_p/2\pi=1\,$MHz, $\Omega_c/2\pi=5.1$\,MHz, and the laser linewidths $\gamma_p/2\pi=0.33\,$MHz and $\gamma_c/2\pi=1.4\,$MHz, typical of recent experiments~\cite{hofmann2012}. The upper red lines correspond to the strong driving case with both Rabi frequencies increased by a factor of $3.75$. In the case of weak driving no excitation enhancement is observed. For strong driving  $f_r/f_0$ exceeds unity (enhancement) and increases towards higher densities (smaller sphere radius). 
Although the laser dephasings are chosen relatively large, the enhancement persists for all densities. This means that the excitation enhancement should be observable under realistic experimental conditions as long as the regime of strong coherent driving ($\Omega_c/\Gamma>\sqrt{\varphi}$) is reached. Through time-dependent simulations we also verified that the steady state is reached in less than $1\,\mu$s for these parameters.  One option to  observe the excitation enhancement effect could  thus be to vary the cloud density at constant atom number, for example by thermal expansion.

The observation of the excitation enhancement has a number of consequences for understanding light-matter interactions in strongly-interacting systems. Firstly, the nonlinear optical response of a Rydberg gas driven under EIT conditions is predicted to obey a universal relation between the optical susceptibility $\chi=\textrm{Im}[\chi_{eg}]$ and the Rydberg fraction \cite{ates2011, sevincli2011b,gaerttner2014b}. This relation states that in the case of perfect EIT $\chi/\chi_{2L} = 1-f_r/f_0$, where $\chi_{2L}$ is the susceptibility in the two-level case, i.e., without coupling laser. Our finding that $f_r>f_0$ would therefore predict a negative susceptibility, hence showing the universal relation must break down in the collectively enhanced regime. Secondly, we show that coherences between $N$-atom collective states and direct two-photon processes are important for determining the steady-state of this system under realistic experimental conditions. This indicates that the classical rate equation (RE) models which neglect these coherences~\cite{ates2011, petrosyan2011, sevincli2011b, heeg2012, hofmann2012, gaerttner2013, gaerttner2014b} are not sufficient to describe the most important details of the experiments. While these models do reproduce the $\sqrt{N}$ collective enhancement, they cannot describe the effect reported here since $f_r<f_0$ is assumed by construction~\cite{gaerttner2014b}. For comparison, the predictions of a RE model~\cite{heeg2012} have been added as dashed lines in Fig.~\ref{fig:fbl_ME}(c), which qualitatively fail to reproduce the enhancement reported here. The excitation enhancement therefore acts as a clear experimental signature for the breakdown of the RE approach.

Finally, we discuss the influence of decoherence. Typically, it is challenging to distinguish between coherent and incoherent excitation dynamics~\cite{schempp2013, schoenleber2013, malossi2013, weber2014, schauss2012, petrosyan2013c, lesanovsky2014, urvoy2014}. The excitation enhancement reported here manifests itself in a global observable in steady-state and is thus comparatively easy to access experimentally. Since it is associated with a direct two-photon process, it requires at least partly coherent dynamics. Indeed, we find that if single atom dephasing is added, $f_r/f_0$ decreases monotonically and falls below unity when dephasing dominates. 

In summary, we have discovered a counter-intuitive excitation enhancement effect that occurs for strong driving of the upper and weak driving of the lower transition of interacting three-level Rydberg atoms. We have shown that this effect is connected to direct multi-photon transitions between collective Dicke states and have derived analytical expressions for the steady state density for arbitrary atom number $N$, which are capable of reproducing the observed enhancement effect in the weak probe limit. The enhancement allows to detect the presence of coherent multi-photon processes, and to identify parameter regimes in which the RE approach breaks down. As the enhancement involves a global steady state observable it should be observable with existing experimental setups \cite{hofmann2012, maxwell2013, peyronel2012, schauss2012}.

A natural next step would be to ask how the observed excitation enhancement is connected to the build-up of multi-partite entanglement. Furthermore, it is interesting to ask whether similar enhancements could be observed in other strongly interacting driven systems, or if the generated multi-particle coherences will find applications in metrology or in quantum information science.

\begin{acknowledgments}
We thank
K.\ P.\ Heeg and M.\ H\"oning
for discussions. 
This work was supported by University of Heidelberg (Center for Quantum Dynamics, LGFG). 
SW acknowledges support by the DFG Emmy Noether grant Wh141-1-1.
\end{acknowledgments}

\clearpage
\newpage

\section{Supplemental material}

Here we provide the details of our analytical calculations of the steady state of a fully blockaded ensemble of atoms in the weak probe limit.

{\it Steady-state perturbation:}
In order to calculate the steady state density matrix perturbatively, we exploit that in the weak probe limit, the ME can be divided into two parts, one of which is proportional to the small parameter $\epsilon=\Omega_p/\Omega_c$:
\begin{equation}
\label{eq:ME_general}
 0=\mathcal{L}_0[\rho]+\epsilon \mathcal{L}_1[\rho] \, .
\end{equation}
We now expand the full steady state in terms of the small parameter $\epsilon$
\begin{equation}
 \rho^{(ss)}=\rho_0+\epsilon \rho_1 + \epsilon^2 \rho_2 + \ldots \,.
\end{equation}
Substituting this into Eq.~\eqref{eq:ME_general} and comparing terms that are of the same order in $\epsilon$ we can solve for the $\rho_i$ recursively assuming that the zeroth order term $\rho_0$, i.e., the steady state of $\mathcal{L}_0$ is known. Then the higher order terms are obtained by solving
\begin{equation}
\label{eq:ss_iterate}
 0=\mathcal{L}_0[\rho_{i}]+\mathcal{L}_1[\rho_{i-1}]
\end{equation}
for $\rho_i$.

In our case we seek to solve
\begin{equation}
\label{eq:ME_full}
 0= -i\frac{\Omega_p}{2}[X_p,\rho] -i\frac{\Omega_c}{2}[X_c,\rho] + \frac{\Gamma}{2}\mathcal{L}_{se}[\rho] \, ,
\end{equation}
where 
\begin{equation}
 X_p=\sum_{i=1}^N\left(\ket{g_i}\bra{e_i}+\ket{e_i}\bra{g_i}\right) \,,
\end{equation}
\begin{equation}
 X_c=\sum_{i=1}^N\left(\ket{e_i}\bra{r_i}+\ket{r_i}\bra{e_i}\right) \,,
\end{equation}
and
\begin{equation}
\label{eq:decay}
 \mathcal{L}_{se}[\rho]=\sum_{i=1}^N\left(2\ket{g_i}\bra{e_i}\rho\ket{e_i}\bra{g_i} - \left\{\ket{e_i}\bra{e_i},\rho\right\}\right) \,.
\end{equation}
The last term accounts for spontaneous decay from the intermediate level.

{\it Symmetrized basis:}
As the Hamiltonian is invariant under exchange of the particles, it is convenient to use a symmetrized basis, or Dicke state basis \cite{dicke1954, lukin2001} (see also \cite{carmele2013}), $\ket{E^j R^s}$ with $s\in\{0,1\}$, $j\in\{0,\ldots, N-s\}$, $\ket{E^0 R^0}=\ket{G}=\ket{g_1\ldots g_N}$ and
\begin{equation}
 \ket{E^j R^s}=\mathcal{N}_{j,s}\left(\sum_{i=1}^N\ket{e_i}\bra{g_i}\right)^j\left(\sum_{i=1}^N\ket{r_i}\bra{g_i}\right)^s\ket{G}
\end{equation}
with the normalization
\begin{equation}
 \mathcal{N}_{j,s}=\sqrt{\frac{(N-j-s)!}{N^s(N-s)!j!}}.
\end{equation}
The matrix elements of the Hamiltonian in this basis are
\begin{equation}
  \begin{split}
    \bra{E^j R^s}X_p\ket{E^{j\prime} R^{s\prime}} & = \sqrt{(j+1)(N-j-s)}\delta_{j,j^\prime - 1}\delta_{s,s^\prime} \\
						      & +\sqrt{j(N-j+1-s)}\delta_{j,j^\prime + 1}\delta_{s,s^\prime}
  \end{split}
\end{equation}
and
\begin{equation}
  \bra{E^j R^s}X_c\ket{E^{j\prime} R^{s\prime}} = \sqrt{j+s}\delta_{j+s,j^\prime + s^\prime}\delta_{1-s,s^\prime}
\end{equation}
We observe that the couplings between the symmetrized states of small $j+s$ induced by the probe laser scale with $\sqrt{N}$, while those of the coupling laser are independent of $N$. This is the reason why all characteristic features in Fig.~3(a) become approximately $N$-independent in the weak probe limit if the parameter $\Omega_p$ is rescaled by $\sqrt{N}$. 

The Lindblad term \eqref{eq:decay} does not preserve the symmetry, so we introduce asymmetric states for $j+s=1$, which are required for the 4th order perturbative calculation
\begin{subequations}
\begin{align}
 \ket{E_k} &=  \frac{1}{\sqrt{N}}\sum_{j=1}^N \ket{e_j}\bra{g_j}{\rm e}^{2\pi i \frac{j k}{N}} \ket{G}  \\
 \ket{R_k} &= \frac{1}{\sqrt{N}}\sum_{j=1}^N \ket{r_j}\bra{g_j}{\rm e}^{2\pi i \frac{j k}{N}} \ket{G} 
\end{align}
\end{subequations}
where for $k=N$ the symmetric states $\ket{E_N}=\ket{E^1R^0}=\ket{E^1}$ and $\ket{R_N}=\ket{E^0R^1}=\ket{R}$ are recovered.

We summarize the action of the Lindblad term on the relevant density matrix elements $\bra{E^j R^s}\rho\ket{E^{j\prime} R^{s\prime}}$. For matrix elements involving a state with no $e$-excitation, we can at most have a diagonal term:
\begin{equation}
 \mathcal{L}_{se}[\ket{E^j R^s}\bra{E^0R^{s\prime}}] = -j\ket{E^j R^s}\bra{E^0R^{s\prime}}
\end{equation}
Matrix elements involving $\ket{E^1}$ decay only to symmetric states:
\begin{equation}
  \begin{split}
 \mathcal{L}_{se}[\ket{E^j R^s} &\bra{E^1}] = -(j+1)\ket{E^j R^s}\bra{G}\\
			   & + 2\sqrt{j\left(1-\frac{j-1+s}{N}\right)}\ket{E^{j-1} R^s}\bra{G} 
 \end{split}
\end{equation}
The action on elements with $j+s=j^\prime+s^\prime=2$ can be summarized as
\begin{equation}
   \begin{split}
 \mathcal{L}_{se}[\ket{E^jR^s} &\bra{E^{j\prime}R^{s\prime}}] = -(j+j^\prime)\ket{E^jR^s}\bra{E^{j\prime}R^{s\prime}} \\
				  &+ 2\sqrt{jj^\prime}\left(1-\frac{1}{N}\right)\ket{E^{j-1}R^s}\bra{E^{j\prime-1}R^{s\prime}} \\
				  &+ \frac{2\sqrt{jj^\prime}}{N(N-1)}\sum_{k=1}^{N-1}\ket{(E^{j-1}R^s)_k}\bra{(E^{j\prime-1}R^{s\prime})_k}
 \end{split}
\end{equation}
with $\ket{(E^1R^0)_k}=\ket{E_k}$ and $\ket{(E^0R^1)_k}=\ket{R_k}$. 

\begin{figure}[t]
  \centering
 \includegraphics[width=\columnwidth]{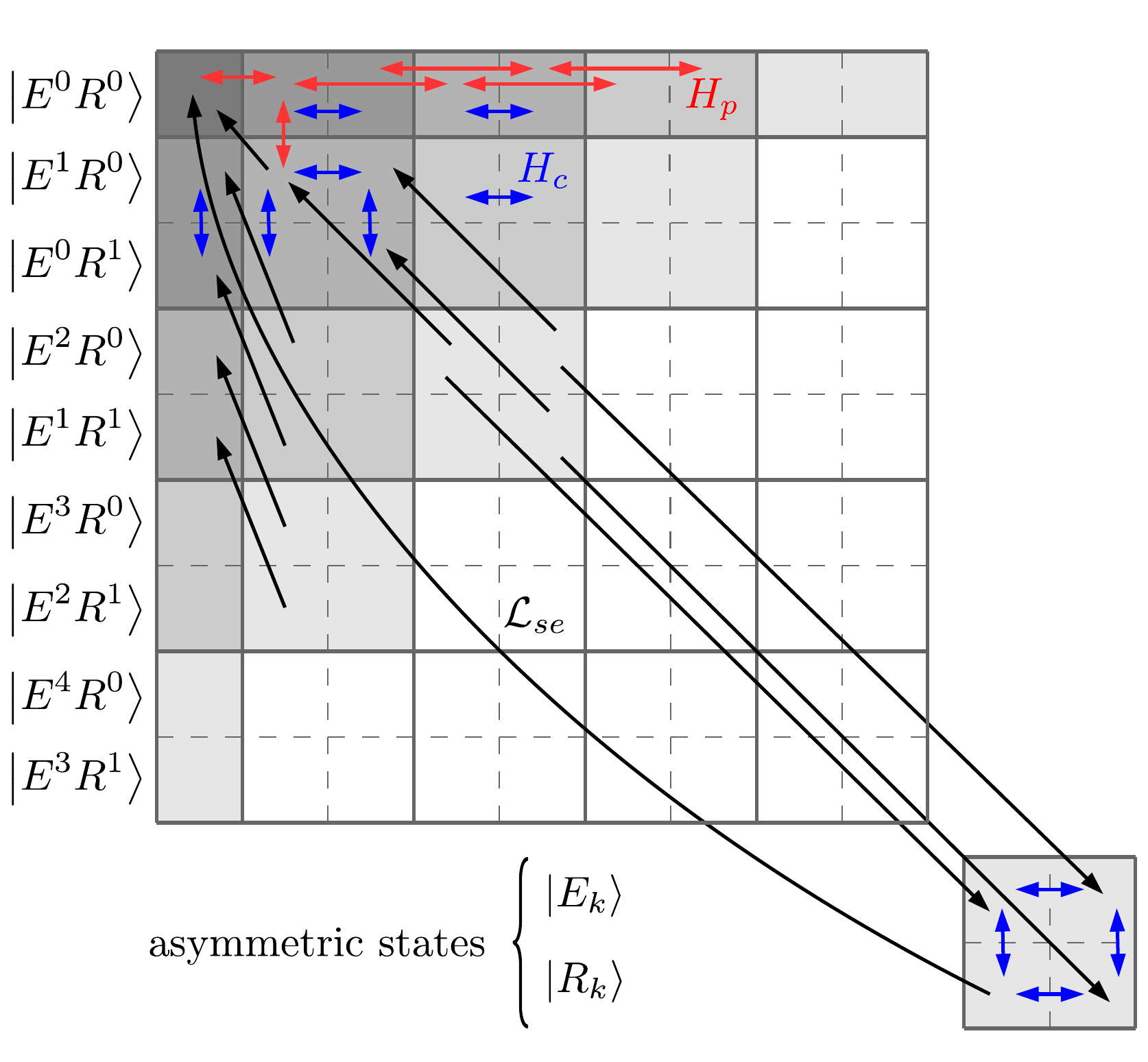}
 \caption{(Color online) Structure of the ME in the symmetrized basis, including couplings between density matrix elements $\ket{E^jR^s} \bra{E^{j\prime}R^{s\prime}}$. $H_c$ connects neighboring cells separated by a dashed line, i.e., differing only in $s$ or $s^\prime$. $H_p$ connects states differing by one in $j$ or $j^\prime$. The Hamiltonian cannot induce couplings between states of different symmetry quantum number $k$. The Lindblad term connects matrix elements in which both $j$ and $j^\prime$ differ by one, irrespective of the symmetry of the state. Note, that only a selection of couplings is shown. The gray shading shows the part of the density matrix that is being populated in the different orders of the weak probe expansion of the steady state, starting with the ground state $\rho_0=\ket{G}\bra{G}$ and extending to higher and higher excited states as the order of $\epsilon$ increases.}
 \label{fig:ME_sketch}
\end{figure}

The structure of the ME is illustrated in Fig.~\ref{fig:ME_sketch}. The time derivative of a matrix element $\rho_{j,s;j\prime,s\prime}$ represented by a cell of the table depends on all the matrix elements from which arrows point to it.

{\it Up to fourth order results in the weak probe limit:}
Here we want to perturbatively solve the steady state equation
\begin{equation}
\label{eq:ME_weakP}
 0= -i\epsilon[X_p,\rho] -i[X_c,\rho] + \beta\mathcal{L}_{se}[\rho] \, ,
\end{equation}
where $\epsilon=\Omega_p/\Omega_c$ and $\beta=\Gamma/\Omega_c$. We have omitted the intermediate state detuning for the sake of readability. In the above notation we thus have $\mathcal{L}_0[\rho]=-i[X_c,\rho] + \beta\mathcal{L}_{se}[\rho]$ and $\mathcal{L}_1[\rho]=-i[X_p,\rho]$.

The zeroth order equation $\mathcal{L}_{0}[\rho_0]=0$ is trivially solved by $\rho_0=\ket{G}\bra{G}$.
Solving the recursion up to the fourth order leads to the unnormalized steady state
\begin{equation}
\label{eq:full_rhoss}
 \begin{split}
  \rho^{(4)}  &= \ket{G}\bra{G} + N\epsilon^2\ket{R}\bra{R} \\
   +  & \frac{\epsilon^4}{x^2} \biggr\{ (N-1)^2\ket{E^1}\bra{E^1}  \\
   & + [N(N-1)(3-\beta^2 x)-(N-1)y]\ket{R}\bra{R} \\
   & + \frac{N(N-1)}{2}\ket{E^2}\bra{E^2} \\
   & + N(N-1)\beta^2\ket{E^1R}\bra{E^1R} \\
   & + \sum_{k=1}^{N-1} \left[x\ket{E_k}\bra{E_k} + y\ket{R_k}\bra{R_k} \right]\biggr\}\\
   & + \text{off-diagonal terms}\,,
 \end{split}
\end{equation}
where we abbreviated $x=1+\beta^2$ and $y=1-\beta^2+\beta^4$.
As we are only interested in populations of the Rydberg and intermediate states, so we do not list all the coherences here. 
From this, one obtains 
\begin{equation}
\label{eq:fr2}
 f_r^{(2)}=\frac{\epsilon^2}{1+ N\epsilon^2}=\epsilon^2 + \mathcal{O}(\epsilon^4)
\end{equation}
and
\begin{equation}
\label{eq:fr4}
 \begin{split}
  f_r^{(4)} & = \frac{1}{\mathrm{Tr}[\rho^{(4)}]}\left[ \epsilon^2 + (N-1)\frac{3-\beta^4}{(1+\beta^2)^2} \right] \epsilon^4 \\
  & = \epsilon^2 + \left[\frac{2(N-1)(1-\beta^2-\beta^4)}{(1+\beta^2)^2}-1\right] \epsilon^4 + \mathcal{O}(\epsilon^6)
 \end{split}
\end{equation}
as well as
\begin{equation}
\label{eq:fe4}
  f_e^{(4)} = \frac{1}{\mathrm{Tr}[\rho^{(4)}]}\frac{2(N-1)}{1+\beta^2}\epsilon^4 = \frac{2(N-1)}{1+\beta^2}\epsilon^4  + \mathcal{O}(\epsilon^6)\,.
\end{equation}
The last expression of Eq.~\eqref{eq:fr4} is what is plotted as a black dotted line in Fig.~1(b) in the main text.
The second order steady state involves at most singly excited states. It is thus equivalent to a three-level atom with $\sqrt{N}$ enhanced probe Rabi frequency and also $f_r^{(2)}$ coincides with the result of a classical rate equation model \cite{heeg2012}.

\end{document}